# Diverse edge states of nanoribbons and excitonic insulator states of the monolayer Ta$_2$Ni$_3$Te$_5$

Hong Tang[*], Jiang Wei[1], Gábor I. Csonka[1], Adrienn Ruzsinszky[†]

*Department of Physics and Engineering Physics, Tulane University*
*New Orleans, LA 70118*

**ABSTRACT**

Ta$_2$Ni$_3$Te$_5$, a layered transition metal chalcogenide with quasi-one-dimensional electronic states, exhibits rich topological and correlated phenomena. Using first-principles calculations, we explore Ta$_2$Ni$_3$Te$_5$ nanoribbons, demonstrating tunable electronic and magnetic properties—ranging from metallic to semimetallic and semiconducting (band gaps of 29.7–60.8 meV), and from ferromagnetic to antiferromagnetic—controlled by edge (Ni or Ta), ribbon width, and H/F saturation. Additionally, GW and Bethe-Salpeter equation (BSE) calculations, complemented by metaGGA-based modified BSE, reveal that the Ta$_2$Ni$_3$Te$_5$ monolayer is an excitonic insulator, with an exciton binding energy exceeding its band gap. These diverse properties position Ta$_2$Ni$_3$Te$_5$ nanoribbons and monolayers as promising candidates for nanoelectronics, spintronics, and optoelectronics, motivating further experimental exploration.

1. **Introduction**

Transition metal chalcogenide (TMC) layered materials are among the main material research nowadays, due to their many novel electronic, magnetic and optical properties [1-6]. TMCs show diverse intralayer atomic structures, including isotropic rotational symmetry [7, 8] and quasi one-dimensional (Q1D) anisotropy [9-12]. The van der Waals (vdW) interaction dominated interlayer interactions allow diverse interlayer stacking and orientation and make the dielectric screening effect in the material substantially reduced and anisotropic, resulting in strong electron-electron [13], electron-hole [14-16], and electron-phonon interactions [12, 17], together with possible diverse band inversions and topology [18, 19].

Recently, a class of TMCs, Ta$_2$M$_3$Te$_5$ (M=Ni and Pd) [11-16, 18, 19], is drawing great attention, showing many exotic quantum states in these materials, including quantum spin Hall effects [18, 20], higher order topological orders [19], superconductivity [13], and Luttinger liquids [11], suggesting interactions between electrons and topology in correlated topological systems.

[*] Email: htang5@tulane.edu
[†] Email: aruzsin@tulane.edu

Additionally, in Ta$_2$Pd$_3$Te$_5$, the evidence of topological exciton insulator state was detected in several experiments [15, 16], showing the rich and complex physics in this material family.

Bulk Ta$_2$M$_3$Te$_5$ are layered orthorhombic crystals with space group Pnma (No. 62). Layers stack via vdW along the cell vector a. In each layer, one Ta atom bonds with five Te atoms to form a TaTe$_5$ tetrahedron. Those tetrahedra connect along vector b via edge-sharing and along vector c via corner-sharing, forming atomic chains along the vector b direction. M (Ni or Pd) atoms occupy the sites near the tetrahedral corners and bases. The anisotropic electron-photon and electron-phonon interactions in Ta$_2$Ni$_3$Te$_5$ [12, 21] and Ta$_2$Pd$_3$Te$_5$ [22] were revealed by polarized angle-resolved Raman spectroscopy. The large anisotropy in conductivity and mobility measured in Ta$_2$Ni$_3$Te$_5$ corroborates the anisotropic electronic structure and electron-phonon interactions [23].

Gou et al. [19] showed the double-band inversion relating to Te-p, Ta-d and Ni-d orbitals in the monolayer Ta$_2$Ni$_3$Te$_5$ with density functional theory (DFT) calculations, revealing the second order topological states with gapped topological edge states and localized corner states, by the Wilson-loop analysis. Double band-inversion commonly happens in Ta$_2$M$_3$Te$_5$ monolayers, making them good platforms for studying topology and interactions in lower dimensions. In gate-voltage modulated Hall bar measurements, Wang et al. [11] detected the power law dependent conductance with temperature and bias voltage in Ta$_2$Pd$_3$Te$_5$ at low temperatures, showing the Luttinger liquid states in quasi 1D conducting channels and suggesting the exotic correlated states by interactions between the band topology and strong correlation in anisotropic low dimensional 2D materials.

Nanoribbons are reduced forms of 2D layered materials, with more spatial confinement effects and rich edge states, making them rife of new functionalities. For example, zigzag MoS$_2$ nanoribbons were calculated as metallic and with ferromagnetic edges [24]. Zigzag WS$_2$ and WSe$_2$ nanoribbons were predicted half-metallic [25]. Also, zigzag graphene nanoribbons can be either ferromagnetic or antiferromagnetic, depending on the ribbon width [26]. The magnetization in CrI$_3$ nanoribbons can be stably aligned either out-of-plane or parallel to the ribbon axis, with more operating controllability [27]. Graphene nanoribbons can support stable and controllable spin centers and chains, topological junction states, and topological end states [28-31]. Also, rich exciton states are shown in SiC [32], MoS$_2$ [33], black phosphorene [34], and CrI$_3$ [27] nanoribbons, and the exciton states can be tuned by bending or strains.

Using first-principles calculations, we investigate the band structures, edge states, and magnetic properties of Ta$_2$Ni$_3$Te$_5$ nanoribbons, exploring the impact of edge (Ni or Ta), ribbon width, and H/F saturation. Our results reveal that these nanoribbons exhibit highly tunable electronic and magnetic properties, ranging from metallic to semimetallic, semiconducting (band gaps of 29.7–60.8 meV), and ferromagnetic to antiferromagnetic states, enabling diverse applications in nanoelectronics and spintronics. We also calculate the electron-hole interaction in the Ta$_2$Ni$_3$Te$_5$ monolayer with many-body perturbation theory GW [35] and BSE (Bethe-Salpeter equation) [36] and metaGGA based modified BSE (mBSE) methods [37, 38]. Our findings demonstrate that the Ta$_2$Ni$_3$Te$_5$ monolayer exhibits an excitonic insulator state, with the lowest exciton's binding energy exceeding its band gap. The versatile electronic, magnetic, and excitonic properties of Ta$_2$Ni$_3$Te$_5$ nanoribbons and monolayers pave the way for innovative applications in nanoelectronics, spintronics, and optoelectronics, motivating further experimental validation and development.

## 2. Computational methods

Nanoribbon calculations are performed with the Vienna Ab initio Software Package (VASP) [39] with projector augmented-wave (PAW) pseudopotentials [40]. The supercell size along the out-of-nanoribbon-plane direction is set to 21 Å to avoid image interaction. The vacuum size between the two edges of periodically repeated nanoribbons is set to 17 Å for narrow ribbons and 36 Å or wide ribbons to avoid the interaction of edges through the vacuum space. The k-mesh of 1x1x20, energy cutoff of 500 eV for plane-wave expansion of wavefunction, and force criterion of 0.01 eV/Å are used for structure relaxation. The self-consistent ground state is calculated with a k-mesh of 1x1x30, and a dense k-mesh of 1x1x120 is for band structure calculations. The r$^2$SCAN metaGGA functional [41] is used. r$^2$SCAN retains metaGGA SCAN's [42] accuracy, with stable numerical computation performance by incorporating the improved interpolation function from rSCAN [43]. Spin-orbit coupling (SOC) is also considered in our calculation.

The G$_0$W$_0$ [35] and G$_0$W$_0$+BSE [36] calculations for monolayer Ta$_2$Ni$_3$Te$_5$ are performed in BerkeleyGW [35] with Quantum ESPRESSO [44]. The starting mean-field (MF) wavefunctions for GW are calculated with PBE (Perdew-Burke-Ernzerhof) [45]. Since SOC has minor effects on the calculated band structure, it is not included in monolayer calculations. The energy cutoff of 90 Ry (1224 eV) and k-mesh of 28x6x1 are for MF wavefunction calculations. The energy cutoff of 20 Ry is for dielectric screening. 1100 bands are included for the band summation and the correction of exact static remainder [46] is used for speeding the convergence with respect to the unoccupied bands. The vacuum layer thickness of 22 Å is inserted, and the Coulomb truncation [35] in this direction is also used to minimize image interactions. A k-mesh of 32x8x1 and both six valence and six conduction bands are set to optical calculations.

The metaGGA based mBSE calculations [37] for excitons are performed with VASP. The ground state wavefunctions are calculated with the r$^2$SCAN and the recent LAK metaGGAs [47]. The dielectric screening is modeled analytically with two screening parameters $\alpha$ and $\mu$, which are obtained by fitting the wavevector-dependent dielectric function calculated from the response function at the level of random phase approximation (RPA). The metaGGA+mBSE method leverages enhanced band gap predictions from advanced metaGGAs (e.g., r$^2$SCAN, LAK) and an accurate screening model, offering superior efficiency for optical calculations compared to traditional approaches like GW+BSE and hybrid+BSE.

## 3. Results and discussion

### 3.1 Theoretical and Experimental Results for the Ta$_2$Ni$_3$Te$_5$ Monolayer

The PBE-optimized Ta$_2$Ni$_3$Te$_5$ monolayer structure, depicted in Figure 1, features cell vector *a* aligned along the atomic chain direction, vector *b* perpendicular to the chains within the 2D plane, and vector *c* out-of-plane (in-plane lattice constants a= 3.717 Å and b=17.796 Å). The r$^2$SCAN and LAK functionals yield similar geometries with slightly different lattice constants (vide infra). The band gap from PBE is 89.5 meV. The gap from PBE+SOC is 89.0 meV. Both are slightly indirect gaps around Γ. The spin-orbit coupling (SOC) effect minimally impacts the calculated band structure, as shown in Figure S1 in SI [48]. The relaxed structure with r$^2$SCAN has lattice constants a=3.713 Å and b=17.709 Å, slightly smaller than those from PBE. The relaxed structure with LAK has lattice constants a=3.760 Å and b=17.973 Å, slightly larger than those from PBE. The band gap from r$^2$SCAN and LAK are 169 and 153 meV, respectively. The band structures from r$^2$SCAN, LAK and r$^2$SCAN+SOC are similar, as shown in Figure S1 in SI [48].

High-quality Ta$_2$Ni$_3$Te$_5$ single crystals were grown using the iodine-assisted chemical vapor transport method. The chemical composition of the crystals was verified by energy-dispersive X-ray spectroscopy (EDS), and their bulk structure was characterized by X-ray diffraction (XRD). Monolayer Ta$_2$Ni$_3$Te$_5$ flakes were prepared by exfoliating the bulk crystals onto SiO$_2$/Si substrates. These flakes were further analyzed using atomic force microscopy (AFM), scanning tunneling microscopy (STM), and high-resolution transmission electron microscopy (HRTEM). Details of the synthesis and characterization are provided in Ref. 21. Scanning tunneling spectroscopy (STS) measurements, conducted under high vacuum and low temperatures, determined the band gap of monolayer Ta$_2$Ni$_3$Te$_5$ to be 90–100 meV, closely matching the PBE value. However, this experimental gap likely represents a renormalized excitonic gap rather than the quasiparticle band gap, due to strong electron-hole interactions characteristic of an excitonic insulator. In Section 3.4, GW+BSE calculations, adjusted via a scissor operation to a 95 meV gap, predict a first dark exciton at -0.195 eV with a binding energy of 285 meV, and metaGGA+mBSE calculations (r$^2$SCAN, LAK) yield binding energies of 490–586 meV, exceeding the band gap (153–169 meV). These large binding energies indicate that the excitonic condensate dominates low-energy excitations, reducing the effective gap probed by STS to the energy of excitonic states (e.g., 0.14–0.21 eV before adjustment, Section 3.4). This behavior is consistent with other excitonic insulators like Ta$_2$Pd$_3$Te$_5$ [15, 16] and 1T-TiSe$_2$ [37], where STS gaps reflect excitonic rather than quasiparticle gaps due to strong electron-hole interactions.

The close agreement between the STS gap (90–100 meV) and the PBE gap (89.5 meV) is likely fortuitous, as PBE underestimates electron correlations but may accidentally approximate the excitonic gap through error cancellation. G0W0@PBE, focusing on quasiparticle states, overestimates the gap (417 meV) by neglecting excitonic effects, while metaGGAs (r$^2$SCAN,

LAK) provide a better starting point for mBSE, which explicitly models electron-hole interactions. Experimental factors, such as tip-sample interactions or substrate-induced screening, may further modulate the STS gap, as the STM tip can induce local fields that couple to excitonic states. These findings highlight the need for combined quasiparticle and excitonic calculations to interpret STS data in excitonic insulators.

## 3.2 Theoretical Results for the Ta$_2$Ni$_3$Te$_5$ Nanoribbons

Fifteen Ta$_2$Ni$_3$Te$_5$ nanoribbons were constructed from the r$^2$SCAN-relaxed monolayer, with their lengths aligned along the atomic chains. Two types of nanoribbons were created: Ni-edged (TNTrb-Ni) and Ta-edged (TNTrb-Ta). For TNTrb-Ni, three widths were built—2TNTrb-Ni, 4TNTrb-Ni, and 6TNTrb-Ni—where the number (2, 4, or 6) indicates the Ta atoms in the unit cell (Figure 2). Each width includes non-saturated nanoribbons and those saturated with four H or F atoms at the edges to passivate dangling bonds, yielding 2TNTrb-Ni-H, 2TNTrb-Ni-F, 4TNTrb-Ni-H, 4TNTrb-Ni-F, 6TNTrb-Ni-H, and 6TNTrb-Ni-F (six passivated nanoribbons). For TNTrb-Ta, two widths were constructed—4TNTrb-Ta and 6TNTrb-Ta—each with non-saturated edges (cf. Figure 3) and edges saturated with two H or F atoms, resulting in 4TNTrb-Ta-H, 4TNTrb-Ta-F, 6TNTrb-Ta-H, and 6TNTrb-Ta-F (four passivated nanoribbons). In total, 15 nanoribbons were studied.

### 3.2.1 Conducting ferromagnetic and antiferromagnetic edge states

Figure 4 presents the calculated properties of the Fluorine-saturated 4TNTrb-Ni-F nanoribbon using R2SCAN with spin-orbit coupling (SOC). The band structure, shown in the left panel of Figure 4(a), reveals a metallic character, with multiple bands crossing the Fermi level. The middle panel of Figure 4(a) displays the density of states (DOS) contributions from edge atoms (four Ni and two Ta, marked by red dots), which dominate the states near the Fermi level, while the right panel shows lesser contributions from non-edge atoms (green dots). The orbital-resolved DOS in Figure 4(e) indicates that Ni d-orbitals primarily contribute to the DOS from -1.0 to 1.0 eV, with Ta d-orbitals contributing more significantly from 0.5 to 1.2 eV. Contributions from Te p-, Ta d-, and F p-orbitals are minimal below 0.0 eV, and other orbitals are negligible within -1.0 to 1.0 eV. This orbital profile resembles that of the Ta$_2$M$_3$Te$_5$ monolayer [19], which has a ~80 meV gap, with valence bands dominated by Ni d-orbitals and conduction bands by Ta d-orbitals due to double band inversion. In 4TNTrb-Ni-F, additional Ni d-orbital contributions appear in the conduction bands near the Fermi level, likely due to edge effects and finite ribbon width, leading to partial (incomplete) band inversion between Ta d- and Ni d-orbitals.

Panels (b) through (d) of Figure 4 depict the spin-resolved band structures, with spin moments projected along the x-axis (ribbon width), z-axis (ribbon thickness), and y-axis (ribbon length), respectively. The spin moments are predominantly along the ribbon thickness (z-axis), with negligible projection along the ribbon length (y-axis) and minimal along the ribbon width (x-axis). Near the Fermi level, up-spin states dominate, indicating a magnetic state. This is corroborated by the magnetization analysis in Figure 4(f), where magnetic moment vectors are shown for each atom (Ta in red, Ni in blue, Te in green, F in grey). The four edge Ni atoms account for nearly all magnetic moments, confirming the nanoribbon's ferromagnetic nature, with arrow sizes proportional to moment magnitude. Figure 4(g) illustrates the SOC energy ($E_{SOC}$) per atom. The four Ta atoms exhibit an $E_{SOC}$ of -0.29 eV, Te atoms -0.20 eV, and Ni and F atoms show smaller values (-0.02 to 0.0 eV). These $E_{SOC}$ magnitudes and trends for Ta, Ni, and Te are comparable to those in the $Ta_2Ni_3Te_5$ monolayer. The large $E_{SOC}$ for Ta atoms arises from their position within a pyramid of five Te atoms, where a strong ligand field enhances Ta d-orbital contributions. Te atoms, located at the pyramid vertices, experience a strong ligand field due to robust Ta-Te bonding, resulting in significant $E_{SOC}$ from p-orbitals. Ni atoms, positioned at vacancy or interstitial sites near the pyramid corners or bases, encounter a weaker electric field, leading to a smaller $E_{SOC}$.

Overall, Figure 4 demonstrates that the 4TNTrb-Ni-F nanoribbon is metallic and edge-ferromagnetic. The significant edge atom contributions to the DOS at the Fermi level, as seen in Figure 4(a), combined with ferromagnetic coupling between the two edges, suggest the potential for spin-polarized currents along the edges, making this nanoribbon a promising candidate for spintronic nanodevices.

The calculated properties of the wider 6TNTrb-Ni-F nanoribbon are presented in Figure S2. This nanoribbon is metallic, with states near the Fermi level predominantly contributed by the six edge Ni and Ta atoms. Orbital DOS analysis reveals similarities to 4TNTrb-Ni-F, with Ni d-orbitals dominating the valence states from -1.0 to 0.0 eV and both Ni d- and Ta d-orbitals contributing significantly to conduction states from 0.0 to 1.2 eV. The two outermost Ni atoms exhibit the largest magnetic moments, antiferromagnetically coupled, with moments primarily aligned along the ribbon thickness direction.

The narrower 2TNTrb-Ni-F nanoribbon shares characteristics with 4TNTrb-Ni-F. It is metallic, with substantial DOS contributions from the six edge Ni and Ta atoms. The orbital DOS trend mirrors that of 4TNTrb-Ni-F, showing significant Ni d- and Ta d-orbital contributions from 0.0 to 1.2 eV and Ni d-orbital dominance from -1.0 to 0.0 eV, driven by incomplete band inversion between Ni d- and Ta d-orbitals. The four edge Ni atoms possess large magnetic moments, ferromagnetically coupled, as depicted in Figure S3 in Supplementary Information.

### 3.2.2 Semiconducting and semimetal nanoribbons

Four hydrogen-saturated nanoribbons—2TNTrb-Ni-H, 4TNTrb-Ni-H, 4TNTrb-Ta-H, and 6TNTrb-Ta-H—are semiconducting, while 6TNTrb-Ni-H is semimetallic. The narrow 2TNTrb-Ni-H exhibits an indirect band gap of 29.7 meV. As shown in Figure 5(a), the two edge Ta and four Ni atoms dominate the states near the Fermi level. The DOS from -1.0 to 1.0 eV, presented in Figure 5(b), is primarily contributed by Ta and Ni d-orbitals, reflecting an incomplete band inversion between these orbitals. The SOC energies, shown in Figure 5(c), are like those in the $Ta_2Ni_3Te_5$ monolayer, owing to the comparable Ta-Te pyramid structure in both systems. Spin moment projections of band states, illustrated in Figures 5(d)–(f), are predominantly along the ribbon width (in-plane polarization), with negligible net magnetization per atom (<0.02 μ_B), indicating no significant magnetic properties.

In contrast, the fluorine-saturated 2TNTrb-Ni-F is metallic and edge-ferromagnetic. The average d-orbital charge on the four edge Ni atoms decreases from 8.39 in 2TNTrb-Ni-H to 8.24 in 2TNTrb-Ni-F. For Ni atoms with more than half-filled d-orbitals, this reduced charge promotes a high-spin state, contributing to the ferromagnetism in 2TNTrb-Ni-F, unlike the non-magnetic 2TNTrb-Ni-H. Additionally, the non-saturated 2TNTrb-Ni is metallic and "p-type" (hole-dominated). Hydrogen saturation in 2TNTrb-Ni-H introduces electron donation, rendering it semiconducting, whereas fluorine in 2TNTrb-Ni-F attracts electrons, enhancing its "p-type" character.

The slightly wider 4TNTrb-Ni-H nanoribbon is semiconducting with a direct band gap of 47.8 meV at the Γ point. Both edge and non-edge atoms contribute to the band states near the Fermi level, as shown in Figure S4(a). The orbital-resolved DOS and SOC energy trends, presented in Figures S4(b) and S4(c), closely resemble those of the narrower 2TNTrb-Ni-H. Like the $Ta_2Ni_3Te_5$ monolayer, 4TNTrb-Ni-H exhibits inversion symmetry, a property shared by (4n)TNTrb-Ni-H nanoribbons for integer n. For sufficiently large n, edge bands may separate from bulk bands, potentially inheriting the topological characteristics of the monolayer's edge bands. The 4TNTrb-Ni-H nanoribbon is non-magnetic, with spin moment projections indicating that band states near the Fermi level are primarily out-of-plane (along the ribbon thickness), as depicted in Figure S4(e). Additionally, two pairs of Dirac-like cones are observed near the Fermi level, with each pair nearly touching at their vertices, as shown in Figure S4(e). These cones, split by spin and offset in energy by approximately 19 meV at the Γ point, resemble the spin-split Dirac cones in monolayer $WSe_2$ [49], though the splitting is significantly smaller (19 meV vs. 0.5 eV in $WSe_2$ [50]). This spin splitting at the band edge influences exciton formation and optical properties.

The widest 6TNTrb-Ni-H, see Figures S5(a)-(f), becomes semimetallic with relatively low DOS around the Fermi level, where the cone touch occurs. Its SOC energy and overall DOS trends are like those of 2TNTrb-Ni-H and 4TNTrb-Ni-H. This nanoribbon is non-magnetic, and its spin projection is mainly along the ribbon width.

The hydrogen saturated nanoribbons 4TNTrb-Ta-H (Figures S6(a)-(f)) and 6TNTrb-Ta-H (Figures S7(a)-(f)) are semiconducting with small direct band gaps of 60.8 and 55.7 meV at Γ, respectively. The band shapes around the vicinity of Γ resemble Dirac cones, showing the approximately linear dispersion of band energy with wavevectors. The orbital resolved DOS of the two nanoribbons shows more weight from Ta d- than Ni d-orbital in the conduction bands within the energy range of 0.0-1.0 eV, indicating a relatively more complete band inversion, compared with above discussed other nanoribbons. The SOC energies of atoms in the two nanoribbons are also like those of monolayer $Ta_2Ni_3Te_5$, echoing a relatively small structure change of the Ta-Te pyramids. The spin projection of 4TNTrb-Ta-H is mainly in the ribbon width direction, while that of 6TNTrb-Ta-H is in both the ribbon width and thickness directions, due to the inversion symmetry in the 6TNTrb-Ta-H nanoribbon. The two nanoribbons are non-magnetic. The DOS around the vicinity of the Fermi level is mainly contributed by Ta and Ni atoms, which form quasi 1D chains in the quasi 1D nanoribbons. The relatively more complete band inversion, small band gap, and quasi 1D features in nanoribbons 4TNTrb-Ta-H and 6TNTrb-Ta-H may make them suitable for observing exotic correlated quantum states, such as the Luttinger liquid, by either thermo-excited or gate-controlled charge carriers [11].

### 3.2.3 Non-magnetic metallic nanoribbons

All the five non-saturated nanoribbons and two Fluorine saturated 4TNTrb-Ta-F and 6TNTrb-Ta-F are metallic and non-magnetic. For the non-saturated 2TNTrb-Ni, the band structure shows a separation at about 0.5 eV above the Fermi level, although those separated bands have a slight overlap in energy, as shown in Figure S8 in SI. Those partially filled bands around the Fermi level make this nanoribbon like "p-type" conducting. By analyzing the d-orbital charge on the four Ni edge atoms, we find that the average d-orbital charge of the edge Ni atoms is 8.36 in the non-saturated 2TNTrb-Ni. This number increases slightly to 8.39 in the H saturated 2TNTrb-Ni-H, however, decreases to 8.24 in the Fluorine saturated 2TNTrb-Ni-F. The different trends on the average d-orbital charge of the edge Ni atoms in those nanoribbons reflect the different chemical behavior from the saturating atoms H and F. H atoms tend to donate electrons to Ni, while F atoms tend to accept electrons from Ni atoms.

The non-saturated 4TNTrb-Ni, see Figure S9 in SI, and 6TNTrb-Ni, see Figure S10 in SI, are qualitatively the same as 2TNTrb-Ni; the Fermi level is crossing the Ni d-orbital derived bands and there is a significant contribution of Ni d-orbital both above and below the Fermi level, making the Ni chains the main conducting channels. The SOC energy trends in the three non-saturated nanoribbons are also like that of monolayer $Ta_2Ni_3Te_5$, indicating the approximately the same structural and chemical environmental situation for the Ta-Te pyramids in those nanoribbons and the monolayer.

The non-saturated 4TNTrb-Ta, see Figure S11 in SI, and 6TNTrb-Ta, see Figure S12 in SI, and the Fluorine saturated 4TNTrb-Ta-F, see Figure S13 in SI, and 6TNTrb-Ta-F, see Figure S14 in SI, nanoribbons share some common features. The orbital resolved DOS shows that around the Fermi level, the contributions from Ta d- and Te p-orbitals noticeably increase, mainly due to the Ta and Te atoms being at the edges in those nanoribbons. For non-saturated 4TNTrb-Ta and 6TNTrb-Ta, the SOC energies of the two outmost edge Te atoms are ~0.25 eV, increased by about 25%, compared to those of other Te atoms in the nanoribbons or in the monolayer. Those outmost Te atoms lonely stick out. The chemical environment around those outmost Te atoms is different from that in the monolayer, where Ni atoms are close to Te atoms both at the vertices and bases of the pyramids. Those lonely sticking-out edge Te atoms are located at the vertices of the pyramids and mostly surrounded by outside vacuum, where the less symmetrical potential environment results in higher SOC energy.

Similarly, for the Fluorine saturated 4TNTrb-Ta-F and 6TNTrb-Ta-F nanoribbons, the F atoms make the edge Ta-Te pyramids significantly deformed. The structural deformation of the edge pyramids influences not only the outmost edge Te atoms, but also the second outmost Te atoms, resulting in a large increase of the SOC energies of the outmost Te atoms and a slight increase of the SOC energies of the second outmost Te atoms.

Those non-magnetic and magnetic conducting nanoribbons present here can be used in the homogenously integrated TMC based electric circuits to be functioned as conducting wires, spin current generators, and memory elements, broadening the design and architecture of novel nanoelectronics.

### 3.4 Excitons in monolayer $Ta_2Ni_3Te_5$

We investigated the optical properties of the $Ta_2Ni_3Te_5$. monolayer, with calculated band structures shown in Figure 6(a). The PBE indirect band gap is ~89 meV, with the valence band maximum (VBM) at Γ and the conduction band minimum (CBM) along the Γ-X line near Γ. In contrast, G0W0@PBE predicts a significantly larger indirect gap of 417 meV, with the CBM at Γ and VBM along the Γ-X line near Γ. This is consistent with the 380 meV gap reported by Dias et al. [51] using the HSE hybrid functional, which includes 25% Hartree-Fock exchange. Both G0W0@PBE and HSE overestimate the experimental STS gap of 90–100 meV, likely due to limitations in capturing strong electron correlations and the double band inversion (Te p-, Ta d-, Ni d-orbitals [19]) characteristic of $Ta_2Ni_3Te_5$. MetaGGAs provide closer agreement with experiment, benefiting from improved self-interaction corrections and local correlation treatment.

Figure 6(b) displays the optical absorption for light polarized along the atomic chain (vector *a*), where the red curve (including electron-hole, e-h, interactions) shows a low-energy peak at ~0.2 eV, compared to ~0.55 eV for the green curve (without e-h interactions), reflecting strong e-h interactions and large exciton binding energies. Figure 6(c) shows absorption for light polarized along vector *b*, revealing distinct peak positions and intensities, indicating anisotropic absorption between the two in-plane directions. Figure 6(d) presents the bound exciton spectrum, featuring a first dark exciton at 0.14 eV (binding energy 276 meV) and a first bright exciton at 0.21 eV (binding energy 211 meV).

The first dark exciton at 0.14 eV arises primarily from the transition between the first valence and first conduction bands near the Γ point. Its wavefunction, shown in Figure 7, is localized in k-space but extends in real space with an in-plane diameter of ~1.5 nm, predominantly along Ta atom chains, reflecting the Ta d-orbital dominance in conduction bands near the Fermi level. The first bright exciton at 0.21 eV results from the transition between the first valence and second conduction bands near Γ, as depicted in Figure 8. It exhibits a similar spatial extent, spreading in the 2D plane as a Wannier-like exciton with a ~1.5 nm diameter. Other excitons display varied spatial extensions, as illustrated in Figures S15 and S16 in Supplementary Information.

To align the band gap with the experimental value (90-100 meV), we applied a scissor operation to adjust the G0W0@PBE gap to 95 meV and performed BSE calculations. The modified quasiparticle band structure is shown in Figure 9(a). Figure 9(b) presents the optical absorption for light polarized along the atomic chains, revealing an anomaly near the gap energy with a negative imaginary macroscopic dielectric function, indicative of exciton states with negative energies, as shown in Figure 9(c). The first dark exciton, at -0.195 eV, has a binding energy of 285 meV, exceeding the 95 meV gap, confirming that the $Ta_2Ni_3Te_5$ monolayer is an excitonic insulator. Figure 9(c) displays multiple dark and bright excitons with negative energies, reflecting strong electron-hole interactions. The first dark (-0.195 eV) and first bright (-0.129 eV) excitons exhibit spatially extended wavefunctions, as illustrated in Figures S17 and S18 in the Supplementary Information, respectively.

We employed the metaGGA+mBSE method to compute the optical properties of the $Ta_2Ni_3Te_5$ monolayer. This approach utilizes band structures and wavefunctions from metaGGAs, incorporating a meta-GGA hybrid exchange-correlation potential ($V_{xc}$) with non-local Fock exchange and a range-separation screening scheme defined by parameters α and μ. These parameters are determined using the CHI method [52] which calculates the static wavevector-dependent dielectric function, followed by fitting to an appropriate screening model (see Figures S19 and S20 in the Supplementary Information). For comparison, we also calculated α using the independent particle dielectric constant via the LOPTICS method [53].

We also employed two metaGGAs, LAK and $r^2$SCAN, to calculate the properties of the $Ta_2Ni_3Te_5$ monolayer. The band structures are shown in Figure S1 in the Supplementary Information, and the first exciton binding energies are listed in Table 1. LAK predicts a band gap of 153 meV, while $r^2$SCAN yields 169 meV. Both functionals use similar screening parameters, resulting in comparable first exciton energies (-420 to -430 meV) using the CHI method. The exciton binding energies, ranging from 580 to 590 meV, exceed the band gap (~100 meV), confirming that $Ta_2Ni_3Te_5$ is an excitonic insulator (EI). For comparison, we calculated the dielectric constant (where α is its inverse) using the LOPTICS method, yielding a smaller α and reduced binding energies (see Table 1), yet still larger than the band gap, reinforcing the EI character.

Figure 10 presents the optical absorption spectra from LAK+mBSE and r2SCAN+mBSE, showing similar profiles with pronounced anisotropy along the two in-plane directions. In bilayer systems, coexisting topological and EI states often feature a spin-orbit gap tied to topological order, heavily influenced by interband hybridization through interlayer tunneling. In contrast, excitons in bilayers, with electrons and holes in separate layers, are bound by long-range Coulomb interactions, minimally affected by interlayer tunneling, potentially causing competition between the quantum spin Hall (QSH) phase and the topologically trivial excitonic phase [54-58]. In monolayers like $Ta_2Ni_3Te_5$, reduced screening amplifies the impact of intralayer interband hybridizations on both the band gap and exciton binding. The gap in $Ta_2Ni_3Te_5$ primarily arises from double band inversion between Ta d- and Te p-orbitals, and Ta d- and Ni d-orbitals, enhanced by quantum confinement, with negligible spin-orbit coupling (SOC) effects [19]. The calculated

Chern number ($Z_2 = 0$) and gapped edge states indicate no first order topological order, but the second order topological order [18]. We propose that the EI states in $Ta_2Ni_3Te_5$ do not alter the $Z_2$ number, as the quasiparticle band structure preserves the band inversion and time-reversal symmetry.

Experimental studies [15, 16] confirm the EI state with zero-momentum excitons in bulk $Ta_2Pd_3Te_5$, evidenced by gap opening detected via polarization-dependent ARPES across various temperatures [15, 16] and surface potassium doping [15, 59]. Topological edge states within the gap formed by the exciton condensate are verified by STS, alongside a second-order EI transition involving non-zero-momentum excitons through STS mapping [16]. Varsano et al. [60] predicted a coexisting quantum spin Hall (QSH) and EI ground state in monolayer T'-$MoS_2$, driven by topological and excitonic orders with high in-plane anisotropy. In contrast, monolayer $Ta_2Ni_3Te_5$ is predicted to host in-gap corner states indicative of second-order topology, based on Wilson loop analysis. The EI state may modulate this higher-order topology via coherent exciton wavefunctions, prompting further theoretical investigation.

4. **Conclusion**

In conclusion, first-principles DFT calculations reveal diverse electronic and magnetic properties of $Ta_2Ni_3Te_5$ nanoribbons, tuned by edge (Ni/Ta), width, and H/F saturation as shown below:

   (i) **Metallic, ferromagnetic and antiferromagnetic nanoribbons:** Fluorine-saturated Ni-edged nanoribbons (2TNTrb-Ni-F, 4TNTrb-Ni-F, 6TNTrb-Ni-F) are metallic, with Ni d-orbitals dominating DOS below the Fermi level and Ni/Ta d-orbitals above, due to incomplete band inversion. Edge Ni atoms couple ferromagnetically in narrower ribbons (2TNTrb-Ni-F, 4TNTrb-Ni-F) and antiferromagnetically in wider 6TNTrb-Ni-F.
   (ii) **Semiconducting, semimetallic and non-magnetic nanoribbons**: Hydrogen-saturated Ni-edged nanoribbons (2TNTrb-Ni-H, 4TNTrb-Ni-H) are semiconducting (gaps ~20–50 meV), while 6TNTrb-Ni-H is semimetallic; Ta-terminated H-saturated nanoribbons (4TNTrb-Ta-H, 6TNTrb-Ta-H) are semiconducting (~60 meV direct gaps).
   (iii) **Metallic and non-magnetic nanoribbons:** Non-saturated (2TNTrb-Ni, 4TNTrb-Ni, 6TNTrb-Ni, 4TNTrb-Ta, 6TNTrb-Ta) and F-saturated Ta-edged nanoribbons (4TNTrb-Ta-F, 6TNTrb-Ta-F). The tunable properties of these nanoribbons enable their use in TMC-based nanoelectronics and spintronics applications.

High-quality $Ta_2Ni_3Te_5$ single crystals were grown using the iodine-assisted chemical vapor transport method. Monolayer $Ta_2Ni_3Te_5$ flakes were prepared and scanning tunneling spectroscopy (STS) measurements determined the band gap to be around 90–100 meV. However, the STS gap may reflect a renormalized excitonic gap, requiring theoretical methods to model both quasiparticle and excitonic contributions accurately. GW and BSE calculations, supplemented by metaGGA-based mBSE, show that the Ta$_2$Ni$_3$Te$_5$ monolayer is an excitonic insulator, with the lowest exciton's binding energy (285–586 meV) exceeding the calculated band gap (95–169 meV). LAK and r$^2$SCAN predict gaps of 153 and 169 meV, respectively, while GW+BSE, LAK+mBSE, and r$^2$SCAN+mBSE confirm strong excitonic binding. Optical absorption spectra exhibit in-plane anisotropy, with excitons involving Ta d- (conduction) and Ni d-orbitals (valence), retaining band inversion and potential topological features. These excitonic and topological properties warrant further exploration.

Acknowledgment:

This work was supported by the donors of ACS Petroleum Research Fund under New Directions Grant 65973-ND10. A.R. served as Principal Investigator on ACS PRF 65973-ND10 that provided support for H.T. A.R. acknowledges support from Tulane University's startup fund. JW acknowledges support from the National Science Foundation (Grant No. 1752997), the Louisiana Board of Regents (Grant No. 082ENH-22), and the Tulane University Carol Lavin Bernick Faculty Grant Program. The computations are carried on the high performance computing (HPC) resources at the Louisiana Optical Network Infrastructure (LONI) and the National Energy Research Scientific Computing Center (NERSC).

Table 1. Calculated results from the LAK and r²SCAN functional related methods for monolayer Ta₂Ni₃Te₅. $E_{gap}$ is the band gap by metaGGA, $E_{exciton}$ is the energy of the first exciton by metaGGA+mBSE, and $E_{binding}$ is the binding energy of this exciton ($E_{binding} = E_{gap} - E_{exciton}$), all with unit of eV. $\alpha$ is the screening parameter (dimensionless) used in the metaGGA+mBSE calculation.

|  | LAK(+mBSE) | | r²SCAN(+mBSE) | |
|---|---|---|---|---|
| $E_{gap}$ | 0.153 | | 0.169 | |
| $\alpha$ | 0.236[a] | 0.108[b] | 0.241[a] | 0.088[b] |
| $E_{exciton}$ | -0.432 | -0.146 | -0.423 | -0.085 |
| $E_{binding}$ | 0.586 | 0.299 | 0.592 | 0.254 |

[a] value from the CHI calculation.
[b] value from the LOPTICS calculation.

Figures

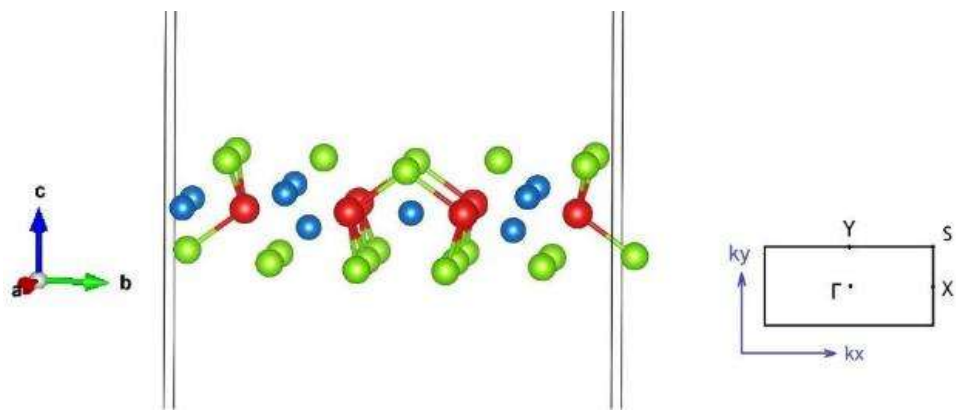

Figure 1. Relaxed atomic structure of the $Ta_2Ni_3Te_5$ monolayer calculated using PBE. The supercell vectors a, b, and c align with the x, y, and z axes of the Cartesian coordinate system, respectively. Ta atoms are shown in red, Ni in blue, and Te in green. The right inset depicts the 2D Brillouin zone in the xy plane.

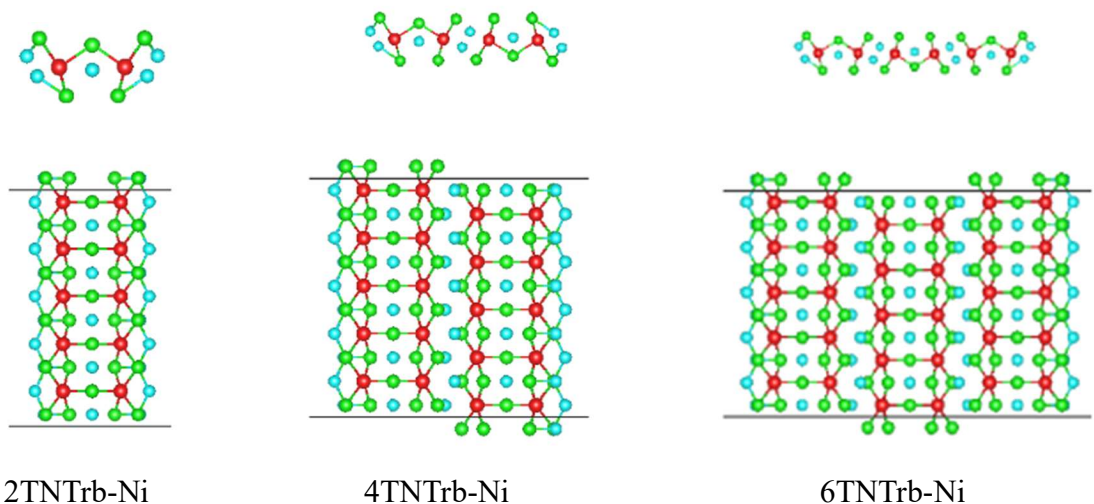

2TNTrb-Ni           4TNTrb-Ni           6TNTrb-Ni

Figure 2. Atomic structures of Ni-edged $Ta_2Ni_3Te_5$ nanoribbons (TNTrb-Ni) with three widths: 2TNTrb-Ni, 4TNTrb-Ni, and 6TNTrb-Ni, where the numbers (2, 4, or 6) denote the Ta atoms in the unit cell. Ta atoms are shown in red, Ni in blue, and Te in green.

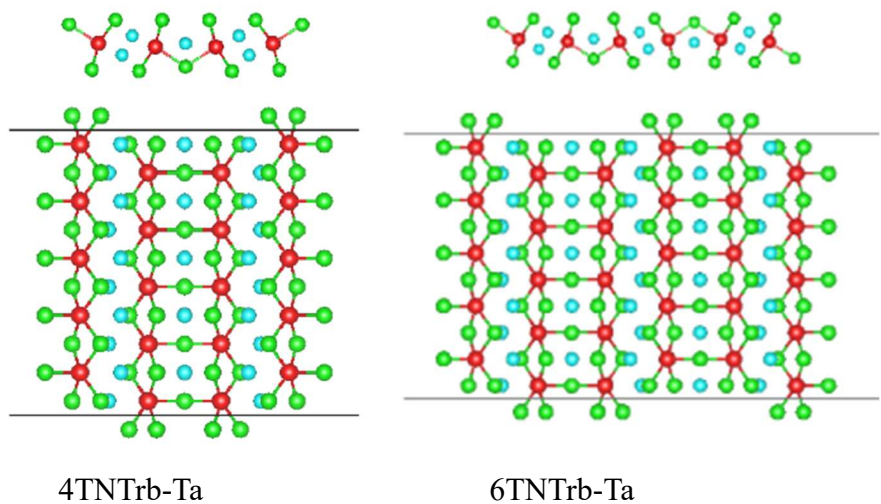

       4TNTrb-Ta                 6TNTrb-Ta

Figure 3. Atomic structures of Ta-edged $Ta_2Ni_3Te_5$ nanoribbons (TNTrb-Ta) with two widths: 4TNTrb-Ta and 6TNTrb-Ta where the numbers (4 or 6) denote the Ta atoms in the unit cell. Ta atoms are shown in red, Ni in blue, and Te in green.

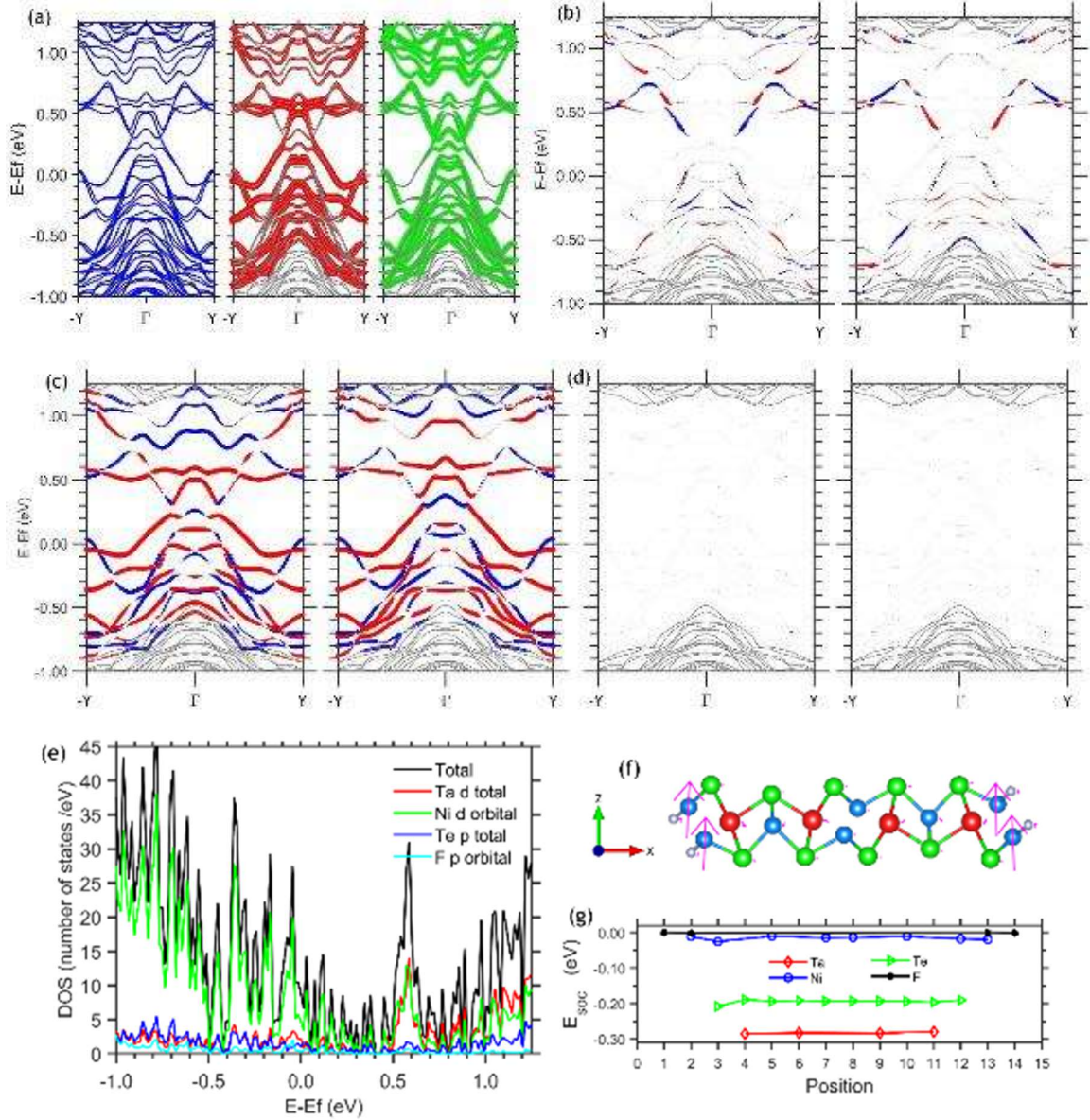

Figure 4. Fluorine-saturated 4TNTrb-Ni-F nanoribbon. (a) Band structure along the atomic chain (vector b, y-axis): left panel shows the total band structure, middle panel highlights DOS contributions from edge atoms (four Ni and two Ta, red dots, as shown in (f)), and right panel shows contributions from non-edge atoms (green dots). Dot size scales with DOS magnitude. (b–d) Spin-resolved band structures with spin moments projected along x (ribbon width), z (ribbon thickness), and y (ribbon length) axes, respectively; odd-indexed bands are on the left, even-indexed on the right for clarity due to likely degeneracy. Red dots represent up-spin, blue dots down-spin, with dot size proportional to magnitude. (e) Partial DOS from different orbitals. (f) Magnetic moment vectors per atom: Ta (red), Ni (blue), Te (green), and F (grey). Moments, except for the four edge Ni atoms, are offset to the right for clarity; arrow size scales with moment magnitude. (g) SOC energy per atom, with atom positions numbered from left to right as in (f).

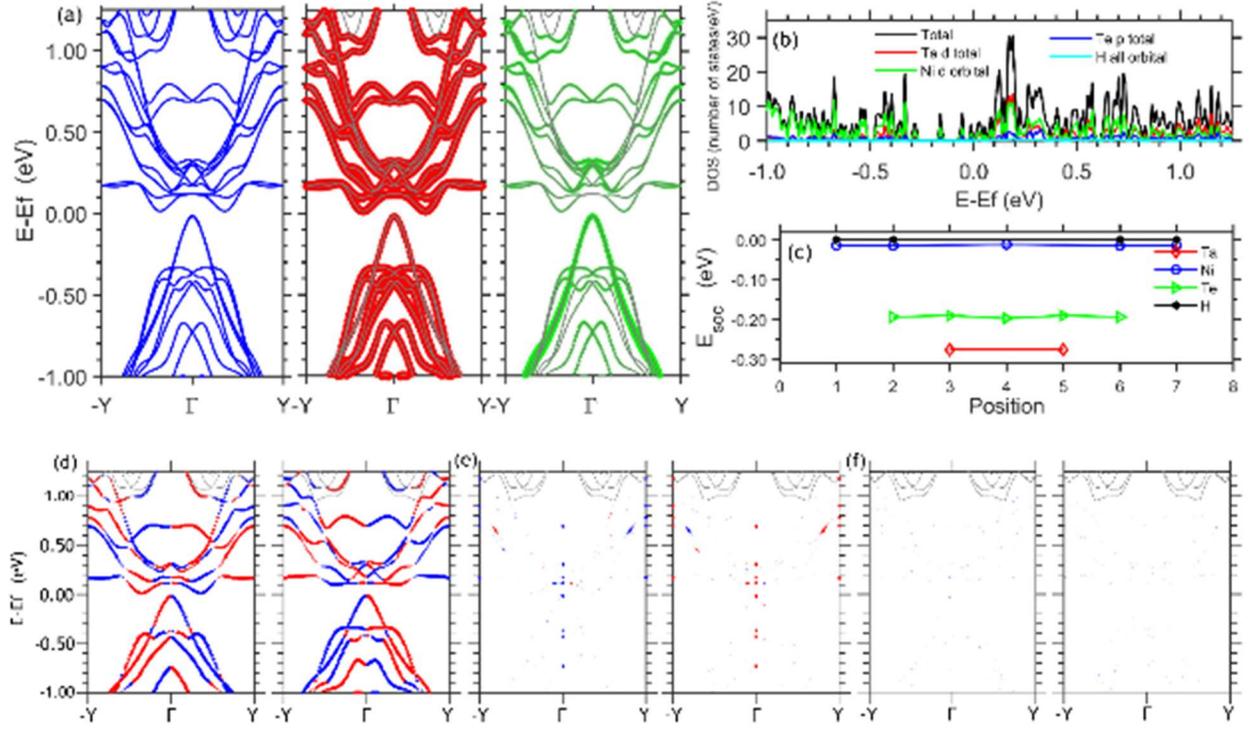

Figure 5. Hydrogen-saturated 2TNTrb-Ni-H nanoribbon. (a) Band structure along the atomic chain: the left panel displays the total band structure, the middle panel shows DOS contributions from edge atoms (red dots), and the right panel shows contributions from non-edge atoms (green dots), with dot sizes proportional to projection magnitude. (b) Partial DOS from different orbitals. (c) SOC energy per atom, with atom positions numbered from left to right along the ribbon width. (d–f) Spin-resolved band structures with spin moments projected along the x (ribbon width), z (ribbon thickness), and y (ribbon length) axes, respectively; odd-indexed bands are on the left, even-indexed on the right for clarity due to likely degeneracy. Up-spin states are marked by red dots, down-spin by blue dots, with dot sizes scaling with magnitude.

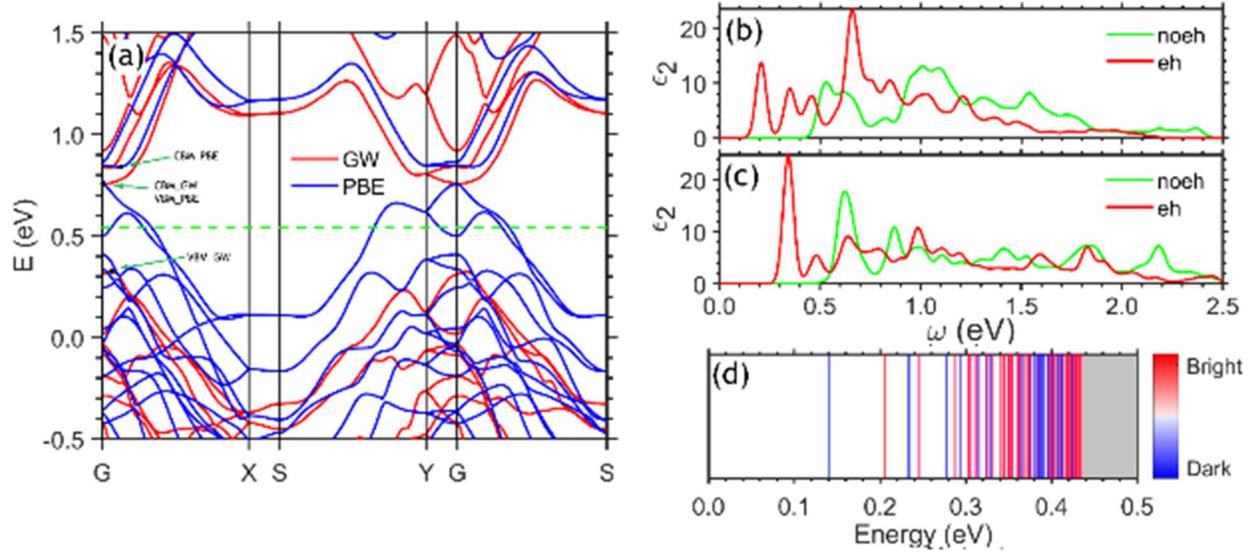

Figure 6. The calculated band structures and optical absorption of monolayer $Ta_2Ni_3Te_5$. (a) shows the calculated band structures by PBE (blue) and $G_0W_0$@PBE (red), obtained via Wannierization without maximal localizations. The dashed green line represents the Fermi level of the $G_0W_0$@PBE calculation. The optical absorption represented by the imaginary part of the macroscopic dielectric function with the light polarized along vector a (or atomic chains) is shown in (b), while the one with the light polarized along vector b is in (c). The red curves are results with the electron-hole (e-h) interaction included, while the green ones are without the e-h interaction. The exciton spectrum is shown in (d), with the red lines for bright excitons and blue lines for dark excitons.

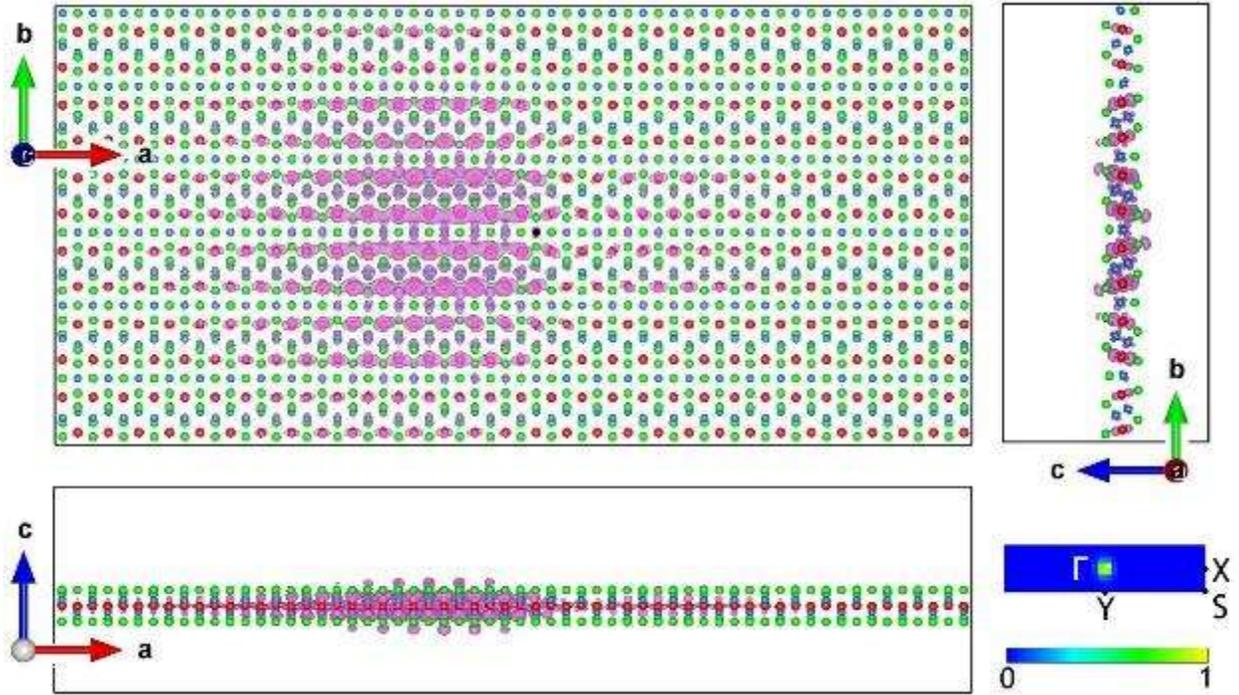

Figure 7. The three isosurface views (in the top-left, top-right and bottom-left subplots) of the squared modulus of wavefunction in real space and the band summed amplitude function $\sum_{v,c}|A_{v,c}(k)|^2$ (the bottom-right subplot) representing the distribution of exciton in k-space, of the first dark exciton at 0.14 eV (shown in Figure 8(d)). The isosurface level is 62.3 $a_0^{-3}$ with $a_0$ being the Bohr radius, and the unit in the color bar of the summed amplitude function is arbitrary.

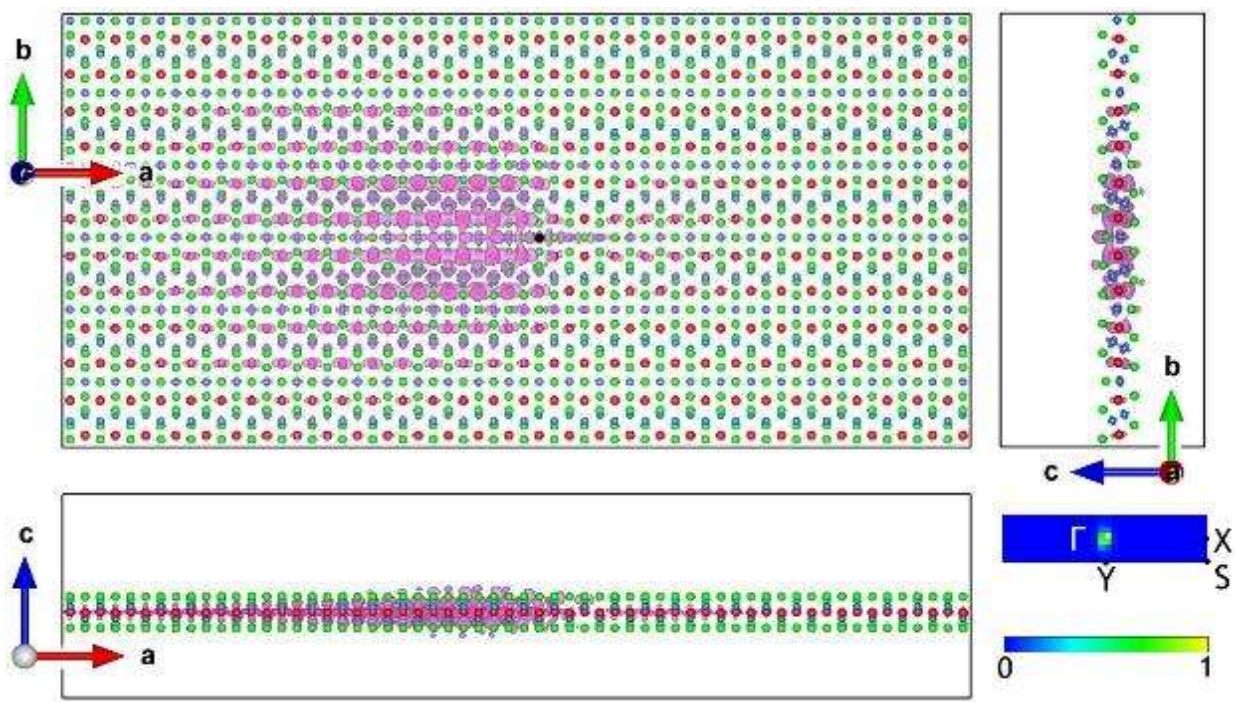

Figure 8. The plots for the first bright exciton at 0.21 eV (shown in Figure 8(d)). The isosurface level is 58.2 $a_0^{-3}$ with $a_0$ being the Bohr radius, and the unit in the color bar of the summed amplitude function is arbitrary.

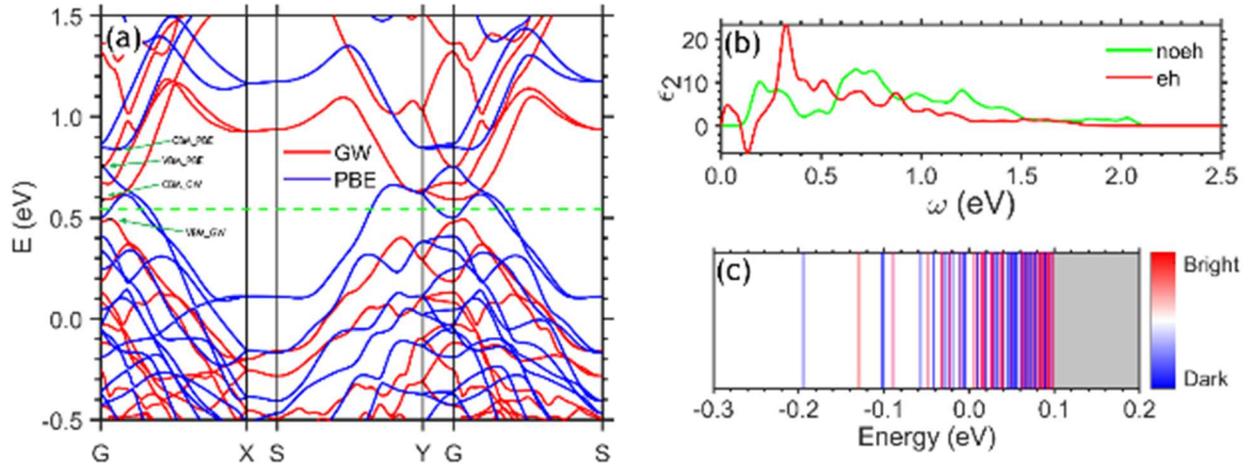

Figure 9. The calculated band structures and optical absorption of monolayer $Ta_2Ni_3Te_5$. (a) shows the calculated band structure by PBE (blue) and the scissor operated band structure of $G_0W_0$@PBE (red). Other subplots are plotted similarly to those in Figure 6.

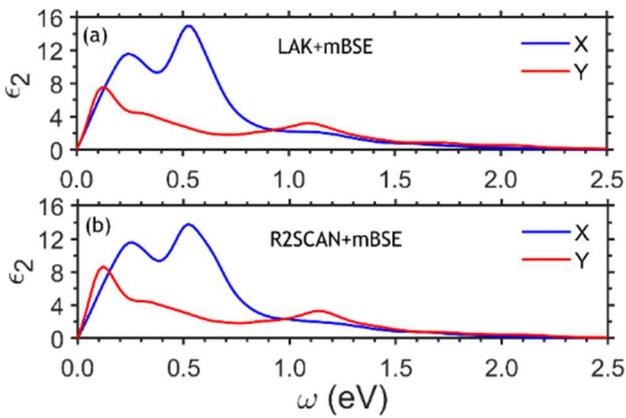

Figure 10. The calculated optical absorption spectrum from LAK+mBSE and R2SCAN+mBSE. The blue curves represent the light polarization along the atomic chain (axis x), while the red curves for the light polarization perpendicular to atomic chains (axis y).